\documentclass[aps,pra,superscriptaddress,amsmath,amssymb,showpacs]{revtex4-2}
\makeatletter\AtBeginDocument{\let\@elt\relax}\makeatother
\usepackage{amssymb,bm}
\usepackage{graphicx}
\usepackage{amsmath}
\usepackage{color}
\usepackage[colorlinks=true,citecolor=blue,urlcolor=blue]{hyperref}
\usepackage{comment}

\begin{document}

\title{Resonant amplification of a relativistic electron bremsstrahlung  on an atom in the undulator field.}
\author{P. A. Krachkov}
\email{p.a.krachkov@inp.nsk.su}
\author{A. I. Milstein}
\email{a.i.milstein@inp.nsk.su}
\author{N. Yu. Muchnoi}
\email{n.yu.muchnoi@inp.nsk.su}
\affiliation{Budker Institute of Nuclear Physics of SB RAS, 630090 Novosibirsk, Russia}
\affiliation{Novosibirsk State University, 630090 Novosibirsk, Russia}

\date{\today}

\begin{abstract}
	It is shown that the influence of the undulator field on the motion of a relativistic electron leads to a significant increase of the bremsstrahlung cross section on an atom near the end of the spectrum of undulator radiation. Observation of this effect is quite a realistic task.
\end{abstract}


\maketitle

\section{Introduction}
At present, the processes of $e^+e^-$ pair production  by a photon in an atomic field and electron bremsstrahlung in this field have been studied in detail using various approaches both at high and low particle energies. At high energies, using the quasiclassical approximation, calculations can be performed exactly in the parameters of an atomic field. A detailed review of the results obtained within this approximation can be found in \cite{KLM16}.

The processes of radiation and $e^+e^-$ pair production in an atomic field are described in terms of cross sections, since the atomic field is localized. Another type of problems includes these processes in macroscopic fields, for example, in laser fields of different spectral composition and polarization. Such processes are described either in terms of probability per unit time or in terms of total probability \cite{Di_Piazza_2012}.

Of great interest is also the study of an influence of macroscopic fields on the processes of quantum electrodynamics in an atomic field. The peculiarity of these problems is that a photon can be emitted by an electron in a macroscopic field without any atomic field. As a result, a singularity arises in the amplitude of the process, which is related to the presence of a cascade process. This singularity cannot be eliminated by introducing a pole into the electron propagator, since it is in the continuous spectrum. Thus, the probability of the process depends on the conditions of the experiment and cannot be described in terms of the cross section.
 
Various results related to the study of an influence of laser fields on bremsstrahlung and pair production processes  in an atomic field can be found in Refs. \cite{Loetstedt_2007, Di_Piazza_2009, Loetstedt_2008,DiPiazza2012,DiPiazza2014, KDM2019,Rosh2019,Rosh2020}. Some effects related to the influence of constant macroscopic fields on processes in atomic fields are described in the book \cite{Baier_1998}. As a rule, the influence of laser fields becomes noticeable at a very high intensity, which complicates the experiments. 
 
 In this paper, we study the bremsstrahlung of a relativistic electron on an atom during the motion of an electron beam in an undulator field. It is shown that the influence of the undulator field will be noticeable at moderate values of the undulator magnetic field and not very high electron energies. In this case, it is possible to choose the region of frequencies and emission angles of the emitted photons in such a way that it will not overlap with the region available for radiation only in the undulator field. As a result, the process can be described in terms of a cross section. However, if the energy  and direction of the emitted photon momentum is close to the accessible region of radiation in the undulator field then  a large increase in radiation (resonant amplification) takes place. We  show that the observation of this effect is a realistic task. All calculations in our work are carried out within the framework of the quasiclassical approximation  exactly in the parameters of the atomic field.

\section{Cross section of the process.}
If a beam of relativistic electrons moves in an undulator, then the cross section $d\sigma$ of a photon radiation by an electron in an atomic field can be represented as the sum $d\sigma=d\sigma_a+d\sigma_f$, where $d\sigma_a$ is the cross section of  radiation in the absence of the undulator field, and $d\sigma_f$ is the contribution related to the undulator field. The bremsstrahlung spectrum $d\sigma_a/d\omega$ and the cross section $d\sigma_a/d\omega d \Omega_{\bm k}$ differential in the photon momentum $\bm k$ have the form
 \cite{BKF}
\begin{align}\label{spectr_small}
&\dfrac{d\sigma_a}{d\omega}=\frac{4\alpha\eta^2}{m^2\omega}\Bigg[\left(1+\frac{\varepsilon'^2}{\varepsilon^2}-\frac{2\varepsilon'}{3\varepsilon}\right)
L+\frac{1}{9}\frac{\varepsilon'}{\varepsilon}\Bigg]\,,\nonumber\\
&\frac{d\sigma_a}{d\omega d\Omega_{\bm k}}=\frac{4\alpha\eta^2\varepsilon^2}{\pi m^4\omega(1+b^2)^2}\Bigg\{[L+\ln(1+b^2)]\left[1+\frac{\varepsilon'^2}{\varepsilon^2}-\frac{\varepsilon'}{\varepsilon}\frac{4b^2}{(1+b^2)^2}\right]\nonumber\\
&-\left[1+\frac{\varepsilon'^2}{\varepsilon^2}+\frac{(1+b^2)^2-10b^2}{(1+b^2)^2}
\,\frac{\varepsilon'}{\varepsilon}\right]
\Bigg\}\,,\nonumber\\
&b= \gamma \theta_{k}\,,\quad \eta=Z \alpha\,, \quad L= \log183 Z^{-1/3}-\mbox{Re}\,\psi(1+i\eta)-C\,,
\end{align}
where $\gamma={\varepsilon}/{m}$,  $Z$ is the atomic charge number, $\alpha$ is the fine-structure constant, $\varepsilon, \varepsilon', \omega$ are the energies of the initial electron, final electron, and photon, respectively, $m$ is the electron mass,  $\theta_{k}$ is the angle between the photon momentum $\bm k$ and the electron momentum $\bm p$,  $\psi(x)=d\Gamma(x)/dx$, $\Gamma(x)$ is the Euler gamma function,  $C=0.577\dots$ is the Euler constant, $\hbar=c=1$.

To calculate the contributions $d\sigma_f/d\omega$ and $d\sigma_f/d\omega d\Omega_{\bm k}$, we use the results of Ref.~\cite{KDM2019}, where  the effect of a laser wave on the bremsstrahlung process of relativistic electron in an atomic field has been studied. In Ref.~\cite{KDM2019}, the laser wave propagated in the direction opposite to the electron momentum $\bm p$. In that case, the vector-potential  of the laser field $\bm A_{l}(t+z)$ can be replaced by $\bm A_{l}(2z)$. The undulator field is described by a similar vector potential $\bm A_{f}(z)$. Therefore, the results for the undulator field can be obtained from the corresponding  results of \cite{KDM2019} for the laser field by replacing $\omega_l\to\omega_0/2$ ($\omega_l$ is the laser frequency) and the dimensionless parameter $\xi$ by the undulator parameter \cite{VL} $K={e B }/{m\omega_0}$, where $e$ is the electron charge, $\bm B$ is the undulator magnetic field, $\omega_0={2\pi}/{\lambda_f}$, $\lambda_f$ is the spatial period of the undulator magnetic field. For a helical undulator, $\bm B$ has the form
$$\bm B(z)=B\Big(\bm e_x\cos(\omega_0 z)+\bm e_y \sin(\omega_0 z)\Big)\,.$$
For $K\ll1$ and $\omega> \omega^*$, where $\omega^*={2\varepsilon^2 \omega_0}/{m^2}$ is the 
end of the spectrum of undulator radiation,  the contribution to the bremsstrahlung spectrum due to the influence of an undulator field reads
\begin{align}\label{spectr_circ}
&\dfrac{d\sigma_f}{d\omega}=\frac{4\alpha\eta^2 K^2}{m^2\omega}\,\Bigg\{\frac{\varepsilon^2+\varepsilon'^2}{\varepsilon^2}\bigg\{
L+\dfrac{1}{2}+\frac{6 L\left(9 y ^2-7\right)-3 y ^2+1}{18\left(y ^2-1\right)^2}+ y^2\bigg[\left(L-\frac{3}{2}\right)l_1\nonumber\\
&-\frac{l_1^2+l_2^2}{8}
-\frac{2 y \left(3 y ^2-2\right) l_2+\left(5-7 y ^2\right) l_1}{12 \left(y ^2-1\right)^2}-\frac{1}{2} \text{Li}_2(\frac{1}{y^2})\bigg]\bigg\}\nonumber\\
&-8y^3\frac{\varepsilon'}{\varepsilon}\bigg[\frac{L\,(y^2-1)}{y^3}+\frac{(12 L+1) \left(7 y ^2-5\right)}{72 y^3 \left(y ^2-1\right)}+\frac{\left(y l_1-l_2\right)}{12 \left(y ^2-1\right)}\nonumber\\
&-\left(L-\frac{3}{2}\right) \left(y l_1+l_2\right)+(y -1) \text{Li}_2\left(\frac{1}{y }\right)+(y +1) \text{Li}_2\left(-\frac{1}{y }\right)\bigg]\Bigg\}\,, \nonumber\\
& y=\omega/\omega^*, \quad l_1=\log (1-1/y ^2)\,, \quad l_2=\log[ (y +1)/(y -1)]\,.
\end{align}
Here $Li_2(x)$ is the dilogarithm. Note that $\omega^*$ is twice less than the maximum radiation frequency in the laser field at $\xi\ll 1$ \cite{KDM2019}. The term \eqref{spectr_circ} contains a singularity at $\omega\to\omega^*$. This is due to the possibility of a cascade process, in which an electron first emits a photon in an undulator field and then elastically scatters in an atomic field, or first scatters in an atomic field and then radiates in an undulator field. Although the cascade process is impossible for $\omega>\omega^*$, closeness of the edge of the undulator radiation spectrum, $\omega-\omega^*\ll \omega$, leads to a significant increase of the bremsstrahlung cross section. At
$1\gg\omega/\omega^*-1\gg 1/(mr_{scr})^2$, where $r_{scr}\sim Z^{-1/3}/m\alpha$ is the screening radius, the contribution to the spectrum has the form
\begin{align}\label{spectr_res}
&\dfrac{d\sigma_f}{d\omega}=\frac{4\alpha\eta^2K^2\omega^*}{3 m^2 (\omega^*-\omega)^2}
\left[L+\frac{1}{2}\log\left(\frac{\omega-\omega^*}{\omega^*}\right)-\frac{1}{6}\right]\,.
\end{align}
Note that $d\sigma_f/{d\omega}$ for $\omega$ close to $\omega^*$ can be noticeably larger than $d\sigma_a/d\omega$. It should be noted  also that the  shape of the spectrum at $0<\omega/\omega^*-1\lesssim 1/(mr_{scr})^2$  is not universal and depends on the macroscopic parameters of the system.

There is also a singularity in the contribution $d\sigma_f/d\omega d \Omega_{\bm k}$ to the differential cross section. This feature is also related to the closeness to the allowed region of undulator radiation. The maximum radiation frequency in the undulator field depends on the photon emission angle:
\begin{equation}
\omega_\theta^*=\dfrac{\omega^*}{1+b^2}\,,\quad  b=\gamma\theta_{k}\,.
\end{equation}
Here $\theta_k$ is the angle between the photon momentum and the undulator axis. At $\omega-\omega^*_\theta\ll\omega$ and $K\ll 1$ we have
\begin{align}
&\frac{d\sigma}{d\omega d \Omega_{\bm k}}=\frac{8\alpha\eta^2K^2\varepsilon^2b^2(1+b^4)\,(\omega_\theta^*)^3}{\pi m^4 (1+b^2)^6(\omega-\omega^*_\theta)^4}\left[ L-5/2+\log(b+1/b)+\log(\omega/\omega^*_\theta-1)\right]\,,
\end{align}
The expression for $d\sigma_f/d\omega d \Omega_{\bm k}$ for arbitrary values of $\omega$ is cumbersome and we do not present it here.

\section {Possibility of experimental observation of the effect.}
Let us now discuss the possibility to observe the effect of a strong increase of the bremsstrahlung cross section on an atom related to the motion in an undulator field. Consider an electron beam with energy $\varepsilon= 430\,\mbox{MeV}$. These is a typical value for the injection complex of the BINP SB RAS \cite{BINP2020}. Let us assume that this beam moves in an axial undulator with length $s = 100\,\mbox{cm}$, spatial period of the magnetic field $\lambda_f =1\,\mbox{cm}$ and magnetic field $B = 0.1\, \mbox{T}$. With such undulator parameters, $\omega^*=175.6\,\mbox{eV}$ and $K=0.093$.
The uncertainty $\omega^*$ associated with the finite length of the undulator is $\delta\omega^*/\omega^*\sim\delta\omega_0/\omega_0=\lambda_f/s =10^{-2}$. The value of $\delta\omega^*/\omega^*$ associated with the energy and angular spread of the electron beam is of the order of $\sim 10^{-3}$. Note that at $\omega>\omega^*$, there are practically no undulator photons, they appear only due to the angular and energy spread of the electron beam, the imperfection of the magnetic system, and the finite number of magnets in the undulator.

\begin{figure}[h]
  	\centering
  	\includegraphics[width=0.6\linewidth]{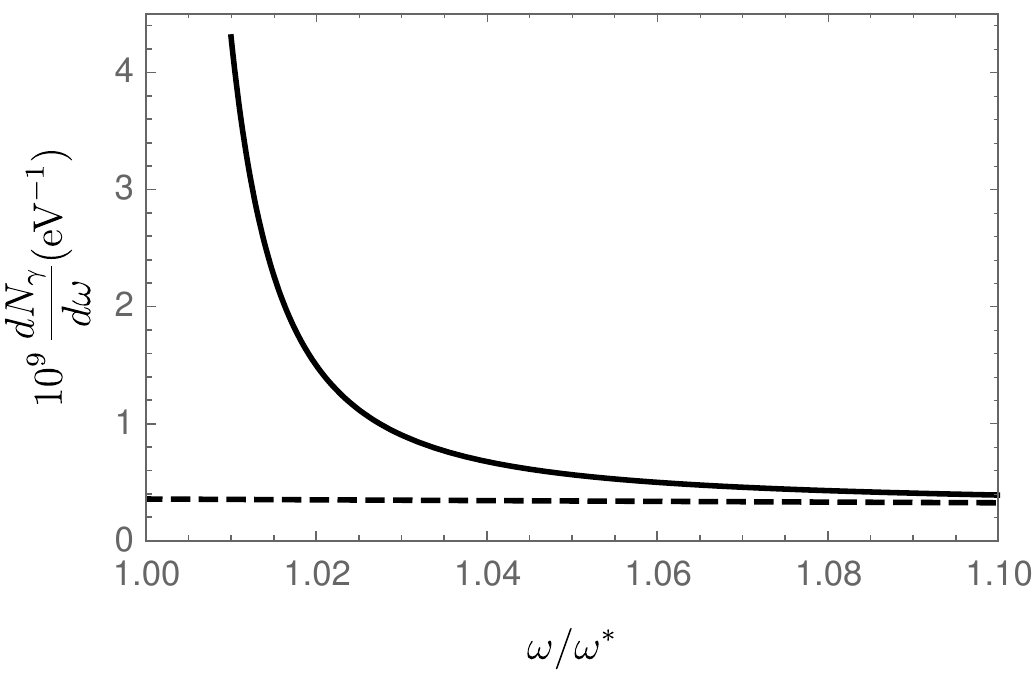}
  	\caption{Dependence of $dN_\gamma/d\omega=ns\,d\sigma/d\omega$ (solid line) and $N^{(0)}_\gamma=ns\,d\sigma_a/d\omega$ (dotted line) on $\omega/\omega^*$ with the parameters specified in the text.}
  	\label{pic}
  \end{figure}

As a target, one can use gaseous helium with density $n=10^{16}\, \mbox{cm}^{-3} $, which fills the  undulator. In Fig.~\ref{pic} the solid line shows the dependence of the photon spectrum $dN_\gamma/d\omega=ns\,d\sigma/d\omega$ on $\omega/\omega^*$ at $1 \gg\omega /\omega^*-1\gg 1/(mr_{scr})^2$. For comparison, the dashed line shows  $N^{(0)}_\gamma=ns\,d\sigma_a/d\omega$. It can be seen that the  undulator field results in significant modification of the photon spectrum near $\omega^*$. The number of photons $N_\gamma$ emitted by one electron in the interval $1.05>\omega/\omega^*>1.01$ is $N_\gamma=8.8\times 10^{-9}$, and in the absence of the undulator $N^ {(0)}_\gamma=2.4\times 10^{-9}$, the difference is almost four times. For  $N=10^{10}$ of electrons in the beam  and the beam injection period $\tau=1\,\mbox{sec}$, we will have $88$ photons per second near the edge of the spectrum of undulator radiation.

\begin{figure}[h]
	\centering
	\includegraphics[width=0.6\linewidth]{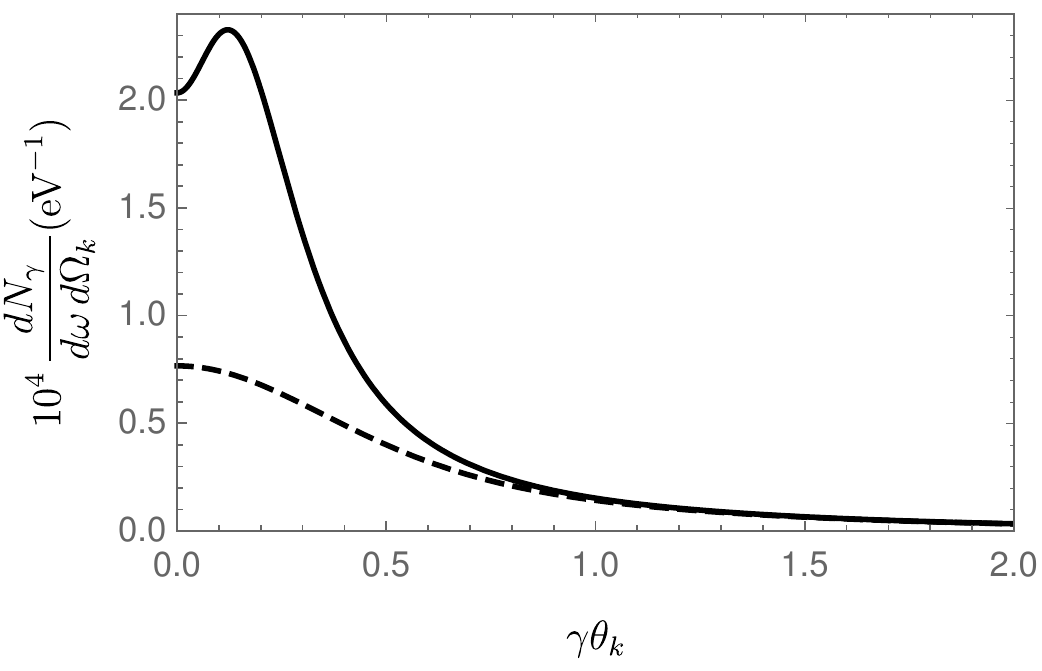}
	\caption{The angular distribution $dN_\gamma/d\omega=ns\,d\sigma/d\omega$ of photons as a function of $\gamma\theta_k$ at $\omega=1.08 \omega^*=189.6\,\mbox{eV}$ (solid line). The dotted line is the angular distribution $dN^{(0)}_\gamma/d\omega\,d\Omega_{\bm k}$.}
	\label{pic1}
\end{figure}

Fig.~ \ref{pic1} shows the angular distribution of bremsstrahlung photons $dN_\gamma/d\omega\,d\Omega_{\bm k}$ (solid line)  near the edge of the spectrum of undulator radiation $\omega^*$ at $\omega=1.08\omega^*=189.6\,\mbox{eV}$. A large peak is seen compared to $dN^{(0)}_\gamma/d\omega\,d\Omega_{\bm k}$ (dotted line). Note that the angle of multiple scattering of electrons when passing through a gas target under these conditions is $10^{-2}/\gamma\ll 1/\gamma$, i.e., multiple scattering does not affect the applicability of the results obtained.

\section{Conclusion}
It is shown that the bremsstrahlung cross section of a relativistic electron moving in a superposition of atomic and undulator fields near the edge of the undulator radiation spectrum is much larger than in the case of the atomic field alone. Above the edge of the undulator spectrum, there are practically no undulator photons; they appear only due to the angular and energy spread of the electron beam, the imperfection of the magnetic system, and the finite number of magnets in the undulator. However, the virtuality of an electron emitting a photon near the edge of the undulator spectrum is very small (the bremsstrahlung photon formation length \cite{BK2004} is large). This is the origin of a significant enhancement of the bremsstrahlung cross section. Note that the use of an undulator instead of laser radiation has an important advantage. In a high-power laser beam, which is necessary to observe the bremsstrahlung amplification effect, the target is destroyed, in contrast to the case of an undulator field. In this paper we have considered the case of a helical undulator. For a planar undulator, the results for a spectrum and angular distribution averaged over the azimuth angle can be obtained   by making the substitution $K^2\to K^2/2$. 

 We show that observation of the interesting coherent effect of bremsstrahlung cross section increasing due to influence of the undulator field is a realistic task.

\end{document}